\documentclass[preprint,aps]{revtex4}

\usepackage{graphicx,subfigure}

\newcommand{\be}{\begin{equation}}
\newcommand{\ee}{\end{equation}}
\newcommand{\bea}{\begin{eqnarray}}
\newcommand{\eea}{\end{eqnarray}}
\newcommand{\bd}{\begin{displaymath}}
\newcommand{\ed}{\end{displaymath}}

\newcommand{\lb }{ \left( }
\newcommand{\rb }{ \right ) }

\newcommand{\da}{D^{\nu}_{t}}
\newcommand{\db}{D^{-\nu}_{t}}

\newcommand{\km}{\frac{k}{m}}

\newcommand{\sn}{\sum_{n=0}^{\infty}}

\begin{document}


\title{
On the fractional damped oscillators and fractional forced oscillators
}

\author{ Won Sang Chung }
\email{mimip4444@hanmail.net}

\affiliation{
Department of Physics and Research Institute of Natural Science, College of Natural Science, Gyeongsang National University, Jinju 660-701, Korea
}

\author{ Min Jung }
\email{sosost1004@nate.com}

\affiliation{
Department of Physics and Research Institute of Natural Science, College of Natural Science, Gyeongsang National University, Jinju 660-701, Korea
}

\date{\today}

\begin{abstract}
In this paper, we use the fractional calculus to discuss the fractional mechanics, where the time derivative is replaced with the fractional derivative of order $\nu$. We deal with the motion of a body in a resisting medium where the retarding force is assumed to be proportional to the fractional velocity which is obtained by acting the fractional derivative on the position. The fractional harmonic oscillator problem, fractional damped oscillator problem and fractional forced oscillator problem are also studied.

\end{abstract}

\maketitle

\section{Introduction}

Fractional calculus implies the calculus of the differentiation and integration whose order is given by a fractional number. The history of the fractional derivatives goes back to the seventeenth century. There are good textbooks for the fractional calculus [1-7].

During about thirty years or so, fractional calculus has attracted much attention due to its application in various fields of science, engineering , finance and optimal problem. For various applications of fractional calculus in physics, see [8-18] and references therein.

Many application of fractional calulus amount to replacing the time derivative in an evolution equation with a derivative of a fractional order. One of the problems encountered in the field is what kind of fractional derivative will replace the integer derivative for a given problem. Non-conservative Lagrangian and Hamiltonian mechanics were investigated by Riewe within fractional calculus [19,20]. Besides, Lagrangian and Hamiltonian fractional sequential mechanics, the models with symmetric fractional derivative were studied in [21,22] and the properties of fractional differential forms were introduced [23].

Two types of fractional derivatives , namely Riemann-Louville and Caputo, are famous. Mathematicians prefer Riemann-Louville fractional derivative because it is amenable to lots of mathematical manipulations. However, the Riemann-Louville fractional derivative of a constant is not zero, and it requires fractional initial conditions which are not generally specified. In contrast, Caputo derivative of a constant is zero, and a fractional differential equation expressed in terms of Caputo fractional derivative requires standard boundary condition. For this reason, physicists and engineers prefer Caputo fractional derivative.

Recently, an extension of the simplest fractional variational problem and the fractional variational problem of Lagrange with constraints was obtained by using Caputo fractional derivative [10]. Even more recently, this approach is extended to Lagrangian formalism [24] with linear velocity. The fractional Hamiltonian formulations were presented for discrete and continuous systems whose dynamics were defined in terms of fractional derivative [25-31].

In this paper we will use the Caputo fractional derivative to discuss the fractional damped oscillator problem and the forced oscillator problem. The Caputo fractional derivative is defined as 
\be
\da f(t)  = \frac{1}{\Gamma (n-\alpha)} \int_0^t ( t- \xi )^{n-\nu -1} \frac{d^n }{d \xi^n }f(\xi) d \xi,
\ee
where $n= [\nu] +1$ and $[x]$ implies a Gauss symbol.

It can be easily checked that the fractional derivative satisfies the following :
\be
\da t^{\beta} = \cases{ \frac{\Gamma ( \beta +1 ) }{\Gamma (\beta - \nu+1 )} t^{ \beta - \nu } &~~~$(\beta \ne 0)$ \cr
0 & ~~~$( \beta = 0 )$ }
\ee

From now on, we will restrict our concern to the case that $ 0< \nu < 1 $. Then the Caputo fractional derivative is given by 
\be
\da f(t) = \frac{1}{\Gamma (1-\nu)} \int_0^t ( t- \xi )^{-\nu } \frac{d }{d \xi }f(\xi) d \xi
\ee

\section {Inverse operator approach}

The simplest first order differential equation is given by
\be
\frac{dN(t)}{dt} = f(t) ~~~(c>0)
\ee
If we consider $ \frac{d}{dt} = D^1_t $, which means the fractional derivative $ D^{\alpha}_t $ with $\alpha =1 $, the eq.(4) is re-written as
\be
D^1_t N(t) = f(t)
\ee
From the following relation for the fractional derivative
\be
D^{\alpha}_t  D^{\beta}_t =  D^{\alpha + \beta}_t, ~~~  D^{0}_t = Id,
\ee
we can solve the eq.(5):
\be
N(t) = D^{-1}_t f(t) = \int_0^t f(\xi ) d \xi ,
\ee
where we insert $\alpha =-1 $ in the eq.(1).

Let us consider the first order fractional kinetic equation
\be
N(t) - N(0) = - c^{\nu}  D^{-\nu}_t N(t),
\ee
which is solved [32] as follows:
\be
N(t) = N(0) E_{\nu} ( - c^{\nu} t^{\nu} )
\ee
Here $E_{\nu} (x) $ is known as Mittag-Leffler function defined by
\be
E_{\nu} ( x) = \sum_{k=0}^{\infty} \frac{x^k}{\Gamma( \nu k +1 ) }
\ee

With a help of the fractional calculus, we are able to construct the classical mechanics related to the fractional calculus. Let us introduce the fractional velocity $v(t)$and fractional acceleration $a(t)$ as follows :
\be
v(t) = \da x(t) , ~~~~ a(t) = \da v(t),
\ee
where $x(t)$ is a fractional position and the time derivative $\da $ is a Caputo's fractional derivative.

In the fractional mechanics, Newton's equation is defined by
\be
F= m a = m \da v ,
\ee
where $m $ is a mass of the body. When the force is constant, the body moves with the constant acceleration $ \frac {F}{m } $.
It should be emphasized that the force is not necessarily constant and, indeed, it may consist of several distinct parts. 
Now let us consider the vertical motion of a body in a resisting medium in which there again exists a retarding force proportional to the fractional velocity. Let us consider that the body is projected downward with zero initial velocity $v(0) =0$ in a uniform gravitational field. The equation of motion is then given by
\be
m \da v = m g -k v
\ee
If we integrate the eq.(13), we obtain
\be
v(t) = g \db (1) - \km \db(v(t))
\ee
If we multiply both sides of the eq.(14) by $ ( -\km )^n D_t^{-n \nu }$ and sum up both sides for $m$ from $0$ to $\infty$, we obtain
\be
\sum_{n=0}^{\infty} ( -\km )^n D_t^{-n \nu }v(t) - \sum_{m=0}^{\infty}  ( -\km )^{n+1} D_t^{-(n+1) \nu }v(t) 
= g \sum_{n=0}^{\infty} ( -\km )^n D_t^{-(n+1) \nu }(1) ,
\ee
which yields
\be
v(t) = \frac{mg}{k} \left[ 1- E_{\nu} \lb - \km t^{\nu} \rb \right]
\ee
Here the asymptotic expansion of $E_{\nu}(-x) $ is given [33] by
\be
E_{\nu}(-x) = \frac{1}{\pi} \sum_{n=0}^{\infty} \frac{ b_n (\nu) }{x^{n+1} }, 
\ee
where $ b_n (\nu) $ is defined by 
\be
 \frac{1}{\pi} \int_{0}^{\infty} dw[ E_{2\nu}(-w^2 ) \cos k w + w E_{2\nu, 1+\nu}(-w^2 ) \sin kw ]
 =  \frac{1}{\pi} \sum_{n=0}^{\infty} b_n (\nu ) k^n 
 \ee
 with
 \be
 b_0 (\nu) =\int_0^{\infty} E_{2\nu} (-t^2 ) dt 
 \ee
and 
  \be
  E_{\alpha, \beta}(x) = \sum_{n=0}^{\infty} \frac{ x^n}{\Gamma( \beta + n\alpha) }
  \ee
  
The function defined by the eq.(20) is a generalization of Mittag-Leffler function and first appeared in the work of Wimam [34].

The asymptotic expansion (17) shows us 
\be
\lim_{t \rightarrow \infty} v(t) = \frac{mg}{k},
\ee
which is a terminal velocity in the fractional mechanics and gives a same result as that in the ordinary mechanics.

Now let us introduce the fractional harmonic oscillator problem :
\be
m D_t^{2\nu} x(t) = - mw^2 x (t), ~~~~ 0<\nu <1
\ee
with the following initial condition
\be
x(0) =A, ~~~~ (\da x )(0) =v_0
\ee
If we integrate the eq.(22), we obtain
\be
\da x(t) - v_0 = - w^2 \db x(t)
\ee
If we integrate the eq.(24) again, we get
\be
x(t) - x(0) = v_0 \db (1) - w^2 D_t^{-2\nu} x(t)
\ee
If we multiply both sides of the eq.(25) by $ ( -w^2 )^m D_t^{-2m \nu }$ , we get
\bea
( -w^2 )^m D_t^{-2m \nu }x(t) - (-w^2) ( -w^2 )^m D_t^{-2(m+1) \nu }x(t) \cr
= A ( -w^2 )^m D_t^{-2m \nu }(1) + v_0 ( -w^2 )^m D_t^{-(2m+1) \nu }(1)
\eea
Now summing up both sides of the eq.(26) for $m$ from $0$ to $\infty$, it gives
\bea
\sum_{m=0}^{\infty} ( -w^2 )^m D_t^{-2m \nu }x(t) - \sum_{m=0}^{\infty}  ( -w^2 )^{m+1} D_t^{-2(m+1) \nu }x(t) \cr
= A \sum_{m=0}^{\infty} ( -w^2 )^m D_t^{-2m \nu }(1) + v_0 \sum_{m=0}^{\infty} ( -w^2 )^m D_t^{-(2m+1) \nu }(1) ,
\eea
which yields
\be
x(t) = A C_{\nu} ( w t^{\nu}) + \frac{v_0}{w} S_{\nu} ( w t^{\nu} )
\ee
and Mittag-Leffler cosine and sine function is defined as
\be
C_{\nu} (x) = \sum_{k=0}^{\infty} \frac{(-1)^k x^{2k}}{\Gamma ( 1 + 2 k \nu )}, ~~~~
S_{\nu} (x) = \sum_{k=0}^{\infty} \frac{(-1)^k x^{2k+1}}{\Gamma ( 1 + (2 k+1) \nu )}
\ee
Mittag-Leffler cosine and sine function are also written as
\be
C_{\nu} (x) = \frac{1}{2} [ E_{\nu} (ix ) + E_{\nu} (-ix) ] , ~~~
S_{\nu} (x) = \frac{1}{2i} [ E_{\nu} (ix ) - E_{\nu} (-ix) ]
\ee

\section{Fractional damped oscillator }

The damped oscillator is the simplest classical model of motion with dissipation which corresponds to friction force proportional to velocity. Analogously, we can consider the fractional damped oscillator defined by
\be
m D_t^{2\nu} x(t) = - mw^2 x (t) - 2 m \gamma \da x(t) , ~~~~ 0<\nu <1
\ee
with the following initial condition
\be
x(0) =A, ~~~~ (\da x )(0) =v_0
\ee

Now we will solve the eq.(31) by using the series approach. Let us assume that the solution of the eq.(31) is given by 
\be
x(t) =\sn c_n t^{n \nu}
\ee 
Inserting the eq.(33) into the eq.(31), we have the following recurrence relation:
\be
a_{n+2} + 2 \gamma a_{n+1} + w^2 a_n =0,
\ee
where
\be
a_n = c_n \Gamma ( 1 + n \nu ) 
\ee

The recurrence relation (34) is re-written as
\be
a_{n+2} - p a_{n+1} = q ( a_{n+1} - p a_n ),
\ee
where $p,q$ are two solutions of the quadratic equation 
\be
t^2 + 2 \gamma t + w^2 =0
\ee
Thus three possibilities exist.

{\bf Case I : over damped oscillator }
In this case $ \gamma^2 >w^2 $ and we put 
\be
p = - \gamma + \sqrt{ \gamma^2 - w^2} , ~~~ q = - \gamma - \sqrt{ \gamma^2 - w^2}
\ee

{\bf Case II : critical damped oscillator }
In this case $ \gamma^2 =w^2 $ and we put
\be
p = q= - \gamma 
\ee

{\bf Case III : under damped oscillator }
In this case $ \gamma^2 < w^2 $ and we put
\be
p = - \gamma + i\sqrt{  w^2 - \gamma^2} , ~~~ q = - \gamma -i\sqrt{  w^2 - \gamma^2}
\ee

Iterating the recurrence relation yields
\be
a_n - p^n a_0 = (a_1 - p a_0) p^{n-1} \sum_{k=0}^{n-1} \lb \frac{q}{p} \rb^k , 
\ee
which gives the following solution
\be
a_n = \cases{ A_1 p^n - B_1 q^n & ~~~$ ( p \ne q )$ \cr
p^n a_0 +  (a_1 - p a_0)n  p^{n-1} &~~~$ ( p = q )$ }
\ee
with
\be
A_1 = a_0 + \frac{ a_1 - p a_0 } {p-q}, ~~~ B_1 =  \frac{ a_1 - p a_0 } {p-q}
\ee 
Thus, the position operator has the following form :
\be
x(t) = \cases{ \frac{v_0 -q A}{p-q} E_{\nu} ( pt^{\nu} ) - \frac{ v_0 -p A}{p-q} E_{\nu} ( q t^{\nu} ) &~~~$ ( p \ne q )$ \cr
A E_{\nu} ( pt^{\nu} ) + \frac{ v_0 - p A }{ \nu } t^{\nu} E_{\nu, \nu} ( p t^{\nu} ) &~~~$ ( p = q )$ }
\ee

In order to discuss the under damped case, we introduce the following functions:
\be
{\tilde C}_{\nu}(a,b| t^{\nu})  = \sum_{m=0}^{\infty}\sum_{n=0}^{\infty}\frac{ (-1)^n \Gamma( 1 + 2n + m ) }{\Gamma( 1 + ( 2n + m ) \nu ) \Gamma( 1 + 2n) \Gamma ( 1 + m ) } a^m b^{2n} t^{(2n +m )\nu} ,
\ee
\be
{\tilde S}_{\nu}(a,b| t^{\nu})  = \sum_{m=0}^{\infty}\sum_{n=0}^{\infty}\frac{ (-1)^n \Gamma( 2 + 2n + m ) }{\Gamma( 1 + ( 2n + m+1 ) \nu ) \Gamma( 2 + 2n) \Gamma ( 1 + m ) } a^m b^{2n+1} t^{(2n +m+1 )\nu} 
\ee   
Two functions satisfy the following relation:
\be
{\tilde C}_{\nu}(a,-b| t^{\nu}) = {\tilde C}_{\nu}(a,b| t^{\nu}), ~~~
{\tilde S}_{\nu}(a,-b| t^{\nu}) = -{\tilde S}_{\nu}(a,b| t^{\nu}), 
\ee
and 
\be
{\tilde C}_{\nu}(a,b| 0) =1 , ~~ {\tilde S}_{\nu}(a,b| 0) =0, ~~~
[\da {\tilde C}_{\nu}(a,b| t^{\nu})]_{t=0} =a , ~~~ [\da {\tilde S}_{\nu}(a,b| t^{\nu})]_{t=0} =b
\ee
Using these function, for the under damped case, the position is given by 
\be
x(t) = A {\tilde C}_{\nu}(-\gamma,w_1| t^{\nu}) + \frac{\gamma A + v_0 }{w_1}{\tilde S}_{\nu}(-\gamma,w_1| t^{\nu}),
\ee
where 
\be
w_1 = \sqrt{ w^2 - \gamma^2 }
\ee
When $\nu $ goes to $1$, we have 
\be
{\tilde C}_{\nu}(a,b| t^{\nu})  \rightarrow e^{at} \cos bt , ~~~
{\tilde S}_{\nu}(a,b| t^{\nu}) \rightarrow e^{at} \sin bt ,
\ee
which implies that the eq.(49) is consistent with the classical result in this limit.

\section{ Forced oscillator}

Now let us consider the forced oscillator with the driven force which is expressed in terms of Mittag-Leffler cosine function. For simplicity, let us  restrict our discussion to the over damped oscillator case. Then the equation of motion is
\be
m D_t^{2\nu} x(t) = - mw^2 x (t) - 2 m \gamma \da x(t) + F_0 C_{\nu} ( \phi t^{\nu} ) , ~~~~ 0<\nu <1
\ee
with the following initial condition
\be
x(0) =A, ~~~~ (\da x )(0) =v_0
\ee

Let us assume that the solution of the eq.(52) is given by
\be
x(t ) =\sn c_n t^{n \nu}
\ee
Inserting the eq.(54) into the eq.(52), we have the following recurrence relation:
\bea
a_{2k+3} + 2 \gamma a_{2k+2} + w^2 a_{2k+1} &=& 0,~~~(k=0,1,2,\cdots)\cr
a_{2k+2} + 2 \gamma a_{2k+1} + w^2 a_{2k} &=& \frac{F_0}{m} (- \phi^2)^k ,~~~(k=0,1,2,\cdots),
\eea
where
\be
a_n = c_n \Gamma ( 1 + n \nu )
\ee

The recurrence relation (55) is re-written as
\bea
a_{2k+3} - p a_{2k+2} &=& q ( a_{2k+2} - p a_{2k+1} ),\cr
a_{2k+2} - p a_{2k+1} &=& q ( a_{2k+1} - p a_{2k} ) + \frac{F_0}{m} (- \phi^2)^k ,
\eea
where  $ \gamma^2 >w^2 $ and 
\be
p = - \gamma + \sqrt{ \gamma^2 - w^2} , ~~~ q = - \gamma - \sqrt{ \gamma^2 - w^2}
\ee
Solving the recurrence relation (57) yields
\bea
a_{2k+1} -p a_{2k} &=& q^{2k} ( a_1-pa_0) + \frac{qF_0}{m} \lb \frac{q^{2k} - (-\phi^2)^k }{q^2 + \phi^2} \rb \cr
a_{2k+2} - p a_{2k+1} &=&  q^{2k+1} ( a_1-pa_0) + \frac{F_0}{m} \lb \frac{q^{2k+2} - (-\phi^2)^{k+1} }{q^2 + \phi^2}\rb
\eea
Inserting the first relation of eq.(59) into the second relation of eq.(59) yields
\be
a_{2k+2} = p^2 a_{2k} + ( p+q) \lb a_1 -p a_0  + \frac{ qF_0 }{m(q^2 +\phi^2)} \rb q^{2k } - \frac{F_0}{m(q^2+\phi^2)} (pq - \phi^2 ) (-\phi^2 )^k
\ee
Solving the eq.(60) and inserting into the eq.(59), we have 
\bea
a_{2k} &=& (a_0 + \eta) p^{2k} + \xi q^{2k} + \zeta (-\phi^2)^k \cr
a_{2k+1} &=& (a_0 p + \eta' ) p^{2k} + (a_1 - p a_0 + \xi' ) q^{2k} + \zeta' (-\phi^2)^k ,
\eea
where
\bd
\eta  =  \frac{1}{p - q }  \lb a_1 -pa_0 + \frac{ qF_0 }{m(q^2 +\phi^2)} \rb - \frac{F_0}{m(q^2 +\phi^2) (p^2 + \phi^2 ) } (q p-\phi^2)
\ed
\bd
\xi  =  \frac{1}{q-p }  \lb a_1 -pa_0 + \frac{ qF_0 }{m(q^2 +\phi^2)} \rb , ~~~ \zeta = \frac{F_0}{m(q^2 +\phi^2) (p^2 + \phi^2 ) } (q p-\phi^2)
\ed
\be
\eta' = p \eta , ~~~ \xi' = p\xi +   \frac{ qF_0 }{m(q^2 +\phi^2)} , ~~~
\zeta' = p\zeta - \frac{ qF_0 }{m(q^2 +\phi^2)}
\ee

Thus, the position has the following form :
\bea
x(t) &=& ( a_0 + \eta) Ch_{\nu} ( pt^{\nu} ) + \xi  Ch_{\nu} ( qt^{\nu} ) + \zeta C_{\nu} ( \phi t^{\nu} ) \cr
&+& (a_0 + \frac{\eta'}{p} ) Sh_{\nu} ( pt^{\nu} ) + \frac{a_1 - pa_0 +\xi'}{q}  Sh_{\nu} ( qt^{\nu} ) + \frac{\zeta'}{\phi} S_{\nu} ( \phi t^{\nu} ) ,
\eea
where $a_0$ and $a_1$ is determined from the initial condition and fractional hyperbolic cosine and sine function is defined as
\be
Ch_{\nu} (x) = \frac{1}{2} [ E_{\nu} (x ) + E_{\nu} (-x) ] , ~~~
Sh_{\nu} (x) = \frac{1}{2} [ E_{\nu} (x ) - E_{\nu} (-x) ]
\ee

\section{Conclusion}
In this paper, we used the fractional Caputo derivative to discuss the fractional mechanics, where the time derivative is replaced with the fractional derivative of order $\nu$. We dealt with the motion of a body in a resisting medium where the retarding force is assumed to be proportional to the fractional velocity which is obtained by acting the fractional derivative on the position. In this case the terminal velocity was shown to be the same as that in the ordinary mechanics. 

We used the inverse operator approach to solve the fractional harmonic oscillator problem. For over damped and critical damped case, the position was shown to be expressed in terms of Mittag-Leffler function. However, for the under damped case, it was shown that the position is expressed in terms of new functions named ${\tilde C}_{\nu}(a,b| t^{\nu}), {\tilde S}_{\nu}(a,b| t^{\nu})$ due to the fact that 
   \bd
   E_{\nu} (x+y) \ne E_{\nu} (x) E_{\nu} (y) . 
   \ed
Finally, we used the series approach to solve the fractional forced oscillator problem with the driven force which is expressed in terms of Mittag-Leffler cosine function.

\def\JMP #1 #2 #3 {J. Math. Phys {\bf#1},\ #2 (#3)}
\def\JP #1 #2 #3 {J. Phys. A {\bf#1},\ #2 (#3)}


\section*{Refernces}

[1] K.Oldham, J.spanier, {\it The fractional Calculus}, Academic Press, New York, 1974.

[2] S. Samko, A. Kilbas, O.Marichev, {\it Fractional Integrals and Derivatives }, Gordon and Breach,   New York, 1993.

[3] K.Miller, B.Ross, {\it An Introduction to the Fractional Calculus and Fractional Differential Equations}, Wiley,  New York, 1993.

[4] A.Kilbas, H.Strivatava, J.Trujillo, {\it Theory and Application of Fractional Differential Equations}, Wiley,  New York, 1993.

[5] I.Pdolubny, {\it Fractional Differential Equations}, (Academic Press) 1999.

[6] R.Hilfer, {\it Application of fractional Calculus in Physics}, World Scientific Publishing Company, Singapore, (2000).

[7] G.Zaslavsky, {\it Hamiltonian Chaos and Fractional Dynamics}, Oxford University Press, Oxford, (2005).

[8] R.Metzler, J.Klafter, J.Phys.A{\bf37}, R161 (2004).

[9] R.Herrmann, Phys.Lett.A {\bf 372} , 5515 (2008).

[10] O.Agrawal, J.Math.Anal.Appl.{\bf 272}, 368 (2002).

[11] R.Almeida, D.Torres, Appl.Math.Lett.{\bf 22} , 1816 (2009).

[12] D.Baleanu, Phys.Scr.{\bf T136} , 014006 (2009).

[13] A.Iomin, Phys.Rev.E {\bf 80}, 022103 (2009).

[14] A.Stanislavsky, Phys.Rev.{\bf E 70} (2004) 051103. 

[15] F.Mainardi, Yu.Luchko and G.Pagnini, Frac.Calc.Appl.Analys. {\bf 4} (2001) 153.

[16] M.Naber, J.Math.Phys. {\bf 45} (2004) 3339.

[17] M.Efe, Asian.J.Control {\bf 14} (2012) 413.

[18] Y.Li, Y.Chen and H.Ahn, Asian.J.Control {\bf 13} (2011) 54.

[19] F.Riewe, Phys.Rev. {\bf E 53} (1996) 1890.

[20] F.Riewe, Phys.Rev. {\bf E 55} (1997) 3582.

[21] M.Klimek , Czech.Journ.Phys. {\bf 52 } (2002) 1247.

[22] M.Klimek , Czech.Journ.Phys. {\bf 51 } (2001) 1348.

[23] C.Shepherd and M.Naber, J.Math.Phys. {\bf 42} (2001) 2203.

[24] D.Baleanu and T.Avkar, Nuovo Cimento {\bf 119} (2004) 73.

[25] E.Rabei and B.Ababneh, arXiv:0704.0519. 

[26] D.Baleanu and T.Avkar, Nuovo Cimento {\bf 119} (2004) 73.

[27] S.Mulish and D.Baleanu, J.Math.anal.Appl. {\bf 304} (2005) 599.

[28] D.Baleanu, Signal Processing {\bf 86} (2006) 2632.

[29] S.Mulish and D.Baleanu, Czech.J.Phys. {\bf 55} (2005) 633.

[30] D.Baleanu and S.Mulish , Physica Scripta {\bf 72} (2005) 119.

[31] D.Baleanu and O.Agrawal, arXiv:0612025v1 (2006).

[32] R.Saxena, A.Mathai, H.Haubold, arXiv:0206240, arXiv:1001.2289.

[33]  M.Berberan-Santos, J.Math.Chem {\bf 38} (2005) 265.

[34] A.Wiman, Acta Mathematica {\bf 29} (1905) 191.

\end{document}